\documentclass[12pt,a4paper,titlepage]{report}
\usepackage{algorithmic}
\usepackage{algorithm}
\usepackage{graphicx}
\usepackage{amsmath}
\usepackage{amssymb}
\usepackage{setspace}
\title{\textbf{Correction of Noisy Sentences using a Monolingual Corpus}}
\author{Thesis submitted in partial fulfillment\\
of the requirements for the degree of\\
Master of Technology\\
in\\
Computer Science and Engineering\\
by\\\\
\textbf{Diptesh Chatterjee}\\
\textbf{Roll No.~06CS3031}\\\\
Under the supervision of\\
\textbf{Dr. Sudeshna Sarkar}\\\\
\includegraphics[height=35mm,width=35mm]{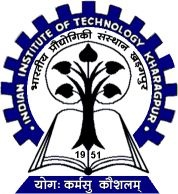}\\
Department of Computer Science and Engineering\\
Indian Institute of Technology, Kharagpur\\
West Bengal, India, 721302
\date {}
}
\begin{document}
\pagestyle{empty}
\maketitle
\newpage
\setlength{\oddsidemargin}{0.50in}
\setlength{\evensidemargin}{0.50in}
\thispagestyle{empty}
\begin{center}
\Large
\textbf{CERTIFICATE} 
\end{center}
\hrule
\hrule
\hrule
\vspace{0.3in}
\begin{flushleft}
This is to certify that the thesis entitled ``\textbf{Correction of Noisy Sentences using a Monolingual Corpus}'' is being submitted by \textbf{Diptesh Chatterjee} (\textbf{Roll No. ~06CS3031}) in partial fulfillment of the requirement for the award of the degree of Master of Technology in \textbf{Computer  Science and Engineering} at Indian Institute of Technology, Kharagpur, India.\\
It is faithful record of work carried out by him at the \textbf{Department of Computer Science and Engineering}, IIT Kharagpur,under my guidance and supervision.
\vspace{1.5in}

\textbf{Prof. Sudeshna Sarkar}\\
\textbf{Department of Computer Science and Engineering}\\
\textbf{IIT Kharagpur}\\
\end{flushleft}
\newpage
\thispagestyle{empty}
\begin{center}
\Large
\textbf{ACKNOWLEDGEMENT}
\end{center}

\hrule
\hrule
\hrule
\vspace{0.3in}

I would like to express my sincere gratitude to Prof.~Sudeshna Sarkar for being my advisor and for providing me the opportunity to conduct research in the field of my interest. I am indebted to her continual support and guidance throughout my project work in the Department of Computer Science and Engineering at IIT Kharagpur.\\
I am really grateful to my friends Arnab Sengupta, Rahul Sarkar, Koushik Hembram and Ashish Jhunjhunwala, who encouraged me at every stage of my project and for being there for me every time the chips were down. I also thank Mr. Sanjay Chatterjee  for providing som data to test my algorithms on.\\
I would also like to avail this opportunity to express my sincere gratitude to my parents and family members who have been an endless source of encouragement for me.\\
Last, but definitely not the least, I would like to thank the developers of C++, Python, Matlab and \LaTeX~for providing me with all the tools required for a successful completion of the project.

\vspace{1in}
{\bf Diptesh Chatterjee}
\newpage

\thispagestyle{empty}
\begin{center}
\Large
\textbf{DECLARATION}
\end{center}

\hrule
\hrule
\hrule
\vspace{0.3in}
I certify that,\\
\begin{enumerate}
\item The work contained in the thesis is original and has been done by myself under the general supervision of my supervisor.
\item The work has not been submitted to any other Institute for any degree or diploma.
\item I have followed the guidelines provided by the Institute in writing the thesis.
\item I have conformed to the norms and guidelines given in the Ethical Code of Conduct of the Institute.
\item Whenever I have used materials (data, theoretical analysis and text) from other sources, I have given due credit to them by citing them in the text of the thesis and giving their details in the reference.
\item Whenever I have quoted written materials from other sources, I have put them under quotation marks and given due credit to the sources by citing them and giving required details in the reference.
\end{enumerate}
\vspace{1in}
{\bf Diptesh Chatterjee}
\newpage

\pagenumbering{roman}
\tableofcontents
\newpage
\thispagestyle{empty}
\listoffigures
\addcontentsline{toc}{chapter}{List of Figures}
\listoftables
\addcontentsline{toc}{chapter}{List of Tables}
\listofalgorithms
\addcontentsline{toc}{chapter}{List of Algorithms}
\newpage
\begin{abstract}
Correction of Noisy Natural Language Text is an important and well studied problem in Natural Language Processing. It has a number of applications in domains like Statistical Machine Translation, Second Language Learning and Natural Language Generation. In this work, we consider some statistical techniques for Text Correction. We define the classes of errors commonly found in text and describe algorithms to correct them. The data has been taken from a poorly trained Machine Translation system. The algorithms use only a language model in the target language in order to correct the sentences. We use phrase based correction methods in both the algorithms. The phrases are replaced and combined to give us the final corrected sentence. We also present the methods to model different kinds of errors, in addition to results of the working of the algorithms on the test set. We show that one of the approaches fail to achieve the desired goal, whereas the other succeeds well. In the end, we analyze the possible reasons for such a trend in performance.
\end{abstract}
\newpage
\pagestyle{plain}
\pagenumbering{arabic}
\onehalfspacing
\chapter{Introduction}
One of the major challenges in Natural Language Processing(NLP) is generation of correct Natural Language sentences. The degree of correctness is measured by how correct the sentence is syntactically and semantically. The branch of NLP dealing with this task is known as Natural Language Generation. However, Natural Language is like an infinite labyrinth of ambiguities, traveling through which does seem to be a rather monumental task. So far, we have no satisfactory computational model of the human brain - one that can explain how humans truly learn, how knowledge is stored and how is it that we can retrieve information extremely fast~\cite{BRAINMODEL}. The algorithms and models used in Natural Language Processing are nothing but collection of mere approximations which perform well on a very select set of data. So, it is quite obvious that such imperfect models will not be able to generate sentences having the quality of actual human-composed sentences.\\\\
The main goal of Natural Language Processing is to allow the human user to interact with the computer using Natural Language. Such imperfections in sentence structure as mentioned previously totally defeat the purpose of NLP. However, due to the absence of actual models of the human brain, at this moment we can only try to improve the quality of the language generated somehow. Such Natural Language Generation is performed by a number of applications like Machine Translation, Question Answering, Dialog Systems, Summarization systems etc. Finding a way to improve the quality of sentences generated by these applications can lead to a better user experience. Also, a large volume of data is generated everyday by non-native speakers of a language. If we have at our disposal methods to improve the quality of noisy sentences, these methods can be used as part of a standalone system in order to \emph{purify} the data generated by the non-native speakers. That is the major motivation behind such corrective mechanisms.\\ \\
Language post-processing is performed as a part of a number of NLP tasks like Machine Translation, Question Answering etc. However, most of the work done is reliant upon the availability of resources in that particular language, which is not always the case. Here, we look into the problem of doing these corrections using only statistical techniques and using minimal language-specific resources. These methods, if successful, would be a great step towards solving the problem of noise-removal in resource-sparse languages.
\section{Language Modeling}
Mathematically, a Language Model~\cite{StatNLP} is expressed as a probability distribution as follows:\\
$P(W_i|W_{i-1}W_{i-2}\ldots W_{i-n+1})$\\
where every $W_j$ is a word in that language. Language models are computed over a corpus of a particular language.It is usually the case that the larger and more diverse the corpus, the language model computed is better. The definition above specifies a $n$-gram language model, where $n$ is known as the order of the model. A language model is in fact, a Markov chain, because the probability that a word occurs is dependent on its preceding words. These probability values are used to score contiguous sequences of words. Given a sequence of $M$ words $S=s_1,s_2,\ldots,s_M$, the score assigned to it by the language model is:\\
$P_{LM}(S)=\prod_j P(s_j|s_{j-1},s_{j-2},\ldots,s_{j-n+1})$\\\\
An important concept in language modeling is that of backoff~\cite{Backoff}. Due to the limitations of the corpus used to compute the language model, it is possible that it might not capture all possible word $n$-grams. So, when a new $n$-gram is encountered, whose probability is not found in the language model, we compute the score using a backoff procedure, which is simply an approximation of the probability in a relaxed setting. For example, one way to backoff might be to try and compute the probability of the word given a smaller history, in effect, retracing to the $(n-1)$-gram and so on. For example:\\
$P(W|W_1,W_2,....W_n)$\\
$\sim P(W|W_1,W_2,....W_{n-1})$    (if $P(W|W_1,W_2,....W_n)$ is not present)\\\\
Another method used to compute scores for unknown $n$-grams is known as smoothing~\cite{smoothing}. In this method, we give the unknown $n$-grams a bit of the probability mass after stripping it off from the known $n$-grams. There are many techniques available for smoothing. The most popular ones are Kneser-Ney discounting, Witten-Bell discounting, and Good-Turing discounting. \\\\
Language model helps capture the fluency of a document, that is, it evaluates how closely a document resembles a piece of Natural Language text. One important thing to note here is that language models do not have anything to do with the semantics of the language. It does not explicitly tell us how correct the sentence is semantically or syntactically, only tells us how closely the sequence of words in the text conforms to the sequence of words in an actual piece of text in that language.
\section{Decoding in Machine Translation}
\begin{figure}
\includegraphics[height=30mm,width=80mm]{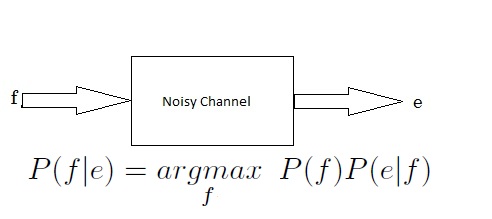}\\
\caption{Noisy Channel Model of Statistical Machine Translation}
\label{fig:NCM}
\end{figure}

Statistical Machine Translation (SMT) is an active field of research in NLP. The fundamental principle used to model SMT is the noisy channel model~\cite{NCM}. Let $e$ be a sentence in the source language. Our aim is to compute the sentence $f$ in the target language which is the best translation for $e$. That is, we wish to select the $f$ which maximizes the probability $P(e|f)$. The Noisy Channel Model assumes that the sentence $e$ and $f$ are essentially the same. The sentence $f$ has been passed through a noisy channel, which has introduced some impurities in the sentence $f$ and transformed it into $e$ as shown in Figure ~\ref{fig:NCM}.\\\\
We model the phenomenon using Baye's rule as follows:\\
\begin{equation}
P(f|e)=argmax_{f}\frac{P(f)P(e|f)}{P(e)}=argmax_{f}P(f)P(e|f)
\label{eq:NCM}
\end{equation}
The denominator term $P(e)$ is removed because it is constant. As we can see from Equation ~\ref{eq:NCM}, the right hand side has two terms. They are defined as follows:
\begin{enumerate}
\item{$P(f)$ is known as the Language Model}
\item{$P(e|f)$ is known as the Translation Model}
\end{enumerate}
The translation model is computed using a parallel corpus~\cite{ParallelCorpus}. A parallel corpus contains aligned pairs of sentences in two languages. The sentences in one language are the translations of the sentences in the other language. A single monolingual corpus is used to compute the language model. The SMT system can be trained on this parallel corpus using techniques such as the IBM models~\cite{NCM} and Minimum Error Rate Training~\cite{MERT}.\\\\
The next task at hand is,given the translation and language models, compute the translation of a sentence in the source language. Usually, SMT systems output the $k$-best list of translations, that is, the best $k$ translations according to the translation and the language models. This process is known as decoding~\cite{NCM}. In phrase-based translation models~\cite{Koehn}, the decoding process involves the following steps:
\begin{enumerate}
\item{Break up the input sentences into a sequence of phrases. Each phrase consists of a contiguous block of words.}
\item{Find the best replacement for each of the phrases according to the phrase table of the translation model.}
\item{Reorder the replaced phrases appropriately using the reordering model which is a part of the translation model.}
\end{enumerate}
Different algorithms have been proposed for decoding. Some of the frequently used decoders are as follows:
\begin{enumerate}
\item{Greedy Decoder~\cite{Greedydec}: It starts with the gloss for words and improves the probability by taking the currently cheapest path. However, it might get stuck in a local minima. To avoid this pitfall, it sometimes uses a 2-step look ahead}
\item{Beam Search based decoder~\cite{pharaoh}: Such a decoder starts translating from left to right and maintains a list of partial hypotheses. When new hypotheses are added to the space, it prunes out the weakest hypotheses in order to accommodate the fixed beam size. But the major problem of such a decoding algorithm is the enormous size of the search space, which can be exponential in the number of hypotheses.}
\item{Word Graph based Decoding~\cite{WordGraphdec}: Such a decoder uses a search graph over the phrases and uses hypothesis recombination to output the $k$-best list of top translations. }
\item{Finite State Transducer Based Decoding: Such decoders model the translation model as a FST with input symbols being the phrases in the source language and output symbols being phrases in target language. The output of the transducer is taken as the translation of the sentence input.}
\item{String to Tree model~\cite{stringtotree}: This is also known as parsing based decoding. It uses dynamic programming, similar in nature to chart parsing. The hypothesis space can be efficiently modeled as a forest structure. A recent work in this area is the use of Synchronous Context Free Grammars (SCFG) in parsing. The hierarchical phrase-based translation model uses such parsing methodology. It also uses a technique called Cube Pruning~\cite{CubePrune} to prune the extremely large hypothesis space. This method has a polynomial time complexity and produces extremely good results.}
\end{enumerate}


\chapter{Previous Work and Problem Definition}
\section{Literature Review}
John Truscott (2007)~\cite{truscott} showed via a series of linguistic experiments that error correction increases a learner's ability to write accurately. The paper refutes previous research stating that the benefits, if any, are very little. He classified the corrective mechanisms in a number of groups and analyzed the factors that probably biased the previous research exdeavours. Hence, correction of noisy text is a problem that is of high practical significance, and, unsurprisingly, has received a lot of attention in the recent years. Out of these, correction of spelling errors is the problem that has probably been the most well studied. The most common approach to solve this problem is maintaining a dictionary of words and computing the replacement candidate by comparing the word to be replaced to each word in the dictionary. But, this would involve a lot of computation. Several methods were proposed to alleviate this problem. For example, Kukich(1992)~\cite{Kukich} , Zobel and Dart(1995)~\cite{Zobel}, De Beuvron and Trinago(1995)~\cite{trinago} proposed similarity keys methods. In these methods, the words in the dictionary are divided into classes according to some word features. The input word is compared to words in classes that have similar features only. \\\\
Another approach to solving word errors in text is finite state automata based approach. Oflazer (1996)~\cite{oflazer} suggested a method where all words in a dictionary are treated as a regular language over an alphabet of letters. All words are represented by a finite state machine atuomaton. For each misspelt word, an exhaustive traversal of the dictionary automaton is initiated using a variant of the Wagner-Fisher algorithm~\cite{wagner} to control the traversal of the dictionary. This method carefully traversals in the dictionary such that inspection of most dictionary states is avoided. Schultz and Mihov (2002)~\cite{schultzmihov} present a variant of this approach where the dictionary is also represented as a finite state automaton. In this technique, a finite word acceptor is constructed foe each word. The acceptor accepts all words that are at a Levenshtein distance~\cite{levenshtein} $k$ from the input word. The dictionary automaton and the Levenshtein automaton are then traversed in parallel to obtain the corrections for each word. Hasan et. al(2008)~\cite{Hassan} propose an approach where they assume that the dictionary is represented as a deterministic finite state automaton. However, they completely avoid computing Levenshtein distance by the use of a Levenshtein transducer. The approach can adopt several constraints on which characters can substitute other characters. These constraints are computed from a phonetic and spatial confusion matrix of characters. Some more recent work include the work in Context Sensitive Spelling Correction, which tries to detect incorrect usage of valid words in a certain context. Using a predefined confusion set is a common approach in this task, for example Golding and Roth (1996)~\cite{rothgoldberg} and Mangu and Brill (1997)~\cite{brillmangu}. Opposite to non-word spelling correction, in this direction, only contextual information was taken into account for modeling by assuming all spelling similarities are equal.\\\\
In the context of Machine Translation, automatic postprocessing has been studied quite widely. But, such methods often target specific phenomena, like correcting English determiners (Knight and Chander, 1994)~\cite{knightchander}, merging German compounds (Stymne, 2009)~\cite{stymne}, or applying word substitution (Elming, 2006)~\cite{elming}. Stymne et. al (2010) ~\cite{stymne2} also proposed another method in using the Swedish grammer checker called Granska to evaluate errors and postprocess Statistical Machine Translation output.\\\\
Reordering of Machine Translation output is another problem that has been studied quite extensively. Xia and McCord (2004)~\cite{xia} describe a method for French, where reordering rules that operate on context-free productions are acquired automatically. Niessen and Ney (2004)~\cite{niessenney} describe an approach for translation from German to English that combines verbs with associated particles, and also reorders questions. Collins et al. (2005)~\cite{collins} also describe an approach for German for reordering of German clauses, which have quite different orders from English clauses. Wang et al. (2007)~\cite{wang} describe a method of Chinese syntactic reordering for SMT systems. Once again, it is a rule based system which specifies reordering rules for different types of phrases in the language and also requires a lot of language specific resource. In fact, this factor is common for all the above mentioned methods. None of them is language independent.\\\\
Gamon et. al (2009)~\cite{gammon} present a system for identifying and correcting English as Second Language (ESL) errors. In this work, they identify the common errors made by native Chinese and Japanese but second language English speakers. They have a number of error specific modules and they are all run in parallel in order to detect errors and correct them. Apart from the standard training data, they have also used the web as a potential source of information. The web is used to offer corrections to text typed in by the user and enable the user to identify whether the text actually matches the user's intent. But the problem with this system is that it is quite time consuming and a person who knows very little of the language cannot possibly benefit from this approach. Also, it was designed specifically for errors made by native Chinese/Japansese speakers. \\\\
Schierle et al. (2007)~\cite{schierle} take spelling correction one step further and use it for text cleaning. They combine the edit distance approach with pointwise mutual information giving the neighbourhood co-occurrence information. They use domain specific replacement strategies and abbreviation lists to increase the correction precision compared to edit distance alone. But the problem with their method is the requirement of a large domain specific corpus, which is readily available for news domain, but not always available.
\section{The Problem Statement}
As mentioned previously, correction of noisy Natural Language sentences has a number of applications. In this work, we try to do the same using a statistical framework. The advantage of using a statistical framework is the fact that we do not need too much of language-specific resources. As a result, we can use the same algorithm in order to solve the problem for different languages.\\\\
The problem we wish to solve is unsupervised correction of noisy Natural Language sentence. To this end, we are given a large monolingual corpus in the language which we are working with. That monolingual corpus can be used as an indicator to the manner in which sentences are constructed in that language. The exact formalism may vary, but we have considered a language model formalism to improve the fluency of the sentences. An important thing to note here is that sentences are characterized by two basic characteristics~\cite{jurafsky}:
\begin{enumerate}
\item{Fluency : Given by the perplexity of a set of sentences according to a language model}
\item{Faithfulness :  Given by one or more metric like the BLEU score~\cite{BLEU}}
\end{enumerate}
In this work, we do not concentrate on the BLEU score. Our aim is only to improve the fluency of the sentences using a language model constructed from the monolingual corpus. Given a noisy sentence and a monolingual corpus, we wish to output the sentence that has the best score according to the language model constructed from the corpus.However, an important thing that we take note of here is the fact that the semantics of the sentence should be kept unaltered as far as possible. For that, we follow a phrase replacement approach. We segment the sentence into a sequence of contiguous phrases and find the best replacements for the phrases that would result in the best score according to the language model. The problem can be mapped to a decoding problem. As previously mentioned, a decoder takes in a phrase translation table, a distortion probability distribution and a language model and a input sentence and after breaking up the sentence into phrases, computes the best substitution for it. Since we are working on a monolingual setting, our phrase translation table is derived from the language model itself. We do not use the distortion probability distribution. Our algorithm works like a decoder with only the language model.\\\\
Given a sentence, finding an optimal segmentation which is in accordance with the cognitive process of human beings is a difficult task. As previously mentioned, this task is an integral part of our problem definition. Several theories have been proposed for this task. However, we do not wish to create a formalism which would be dependent on any kind of external language-specific resource. As a result, we do not follow any rigorous linguistic method of phrase segmentation. We define a phrase as a contiguous sequence of words in a sentence and proceed to replace individual phrases in accordance with this definition. The next step is combination of the replaced phrases so that the final sentence has the highest possible score according to the language model. We aim to achieve an output sentence which has a more natural word order than the original sentence, but we also aim to keep the meaning preserved as much as possible.\\\\
In order to find replacement candidates for phrases, we need to model the errors.  Some of the commonly observed errors in written English are~\cite{error}:
\begin{enumerate}
\item Singular/Plural Form
\item Verb Tense
\item Word choice
\item Subject/Verb Agreement
\item Preposition
\item Spelling errors
\item Word form
\item Redundancy
\item Missing word
\item SVO/SOV word orders for non-native speakers of language
\end{enumerate}
We can broadly classify these errors into the following categories:
\begin{enumerate}
\item Reordering based errors
\item Substitution based errors
\item Spelling errors
\item Missing words
\end{enumerate}
The phrase replacement selection function should be such that it is able to accommodate all the above mentioned errors, and also be able to preserve the semantics of the original sentence. As a result, it is necessary to select a phrase that is not very different from the original phrase in terms of content. Also, all the above mentioned errors have to be modeled using a number of distance functions. The final distance function should be a combination of components for modeling all the above errors. Each component of the distance function computes a similarity score between the phrase to be replaced and its possible replacement in accordance with the type of error which that component is designed to model. The final score is the combination of all the individual scores.\\\\
Since noisy data can come from various sources, the errors that occur in such data are varied. For example, in chat data, we can see a number of abbreviated forms or non-standard language being used. In blogs, we see a number of out of dictionary words being used. Such data more often than not do not adhere to the standard syntax. Machine translation systems is another way of obtaining noisy data. A major problem with such data is the incorrect usage of synonyms resulting from a lack of good bitext. Also, if the machine translation system does not have a good enough reordering model, we can see a number of reordering based problems that can arise if the source and target languages have different word orders. So, the distance function being used is dependent upon the application and should be tuned accordingly.\\\\
\section{Evaluation Metrics}
The main purpose of this work is to improve the correctness of a sentence given a Language Model. Since a Language Model helps measure the fluency of a piece of text, the main quantity being measured here is the \textbf{fluency}. Although fluency is not a sufficient metric to measure how good a piece of text is. This is because it is possible to string together a set of meaningful chunks in order to form a sentence. Such a sentence will get a good score according to the Language Model, but will not count as a good sentence in the language. As a result, we should also measure the \textbf{faithfulness} of the text. Faithfulness can be quantified by how close a given piece of text is in comparison to a human written piece of text. This is done by comparing the machine-generated text to a piece of standard reference human-written text.\\\\
In order to quantitatively measure the correctness of sentences, we need numerical methods in order to judge the fluency and faithfulness of the sentences generated. Fluency is measured by the Perplexity score according to a language model and the faithfulness is measured by the BLEU score. 
\subsection{Perplexity}
Perplexity helps to determine how probable a given sentence is according to a Language Model. In other words, it decides how the phrases will be ordered by a fluent speaker of the language. The ordering of phrases is the key factor here, as perplexity does not take into consideration any kind of syntactic or semantic rules that might govern the formation of a correct sentence in any language. Perplexity is purely an Information Theoretic measure.\\\\
Say, we are given a language model $LM$ which defines a probability distribution $P_{LM}$. Let us say that $LM$ is an $N$-gram language model. Say we have a sentence $S=S_1,S_2,\ldots , S_l$ of $l$ words. We define the \textbf{average logprob} per word as follows:\\\\
Average Logprob ($LP$) = $-\frac{1}{l}\sum_{i=N}^{l}log P_{LM}(S_i|S_{i-N+1}S_{i-N}\ldots S_{i-1})$\\\\
The average perplexity per word is defined as :\\\\
Perplexity ($PP$) = $2^{LP}$\\\\
We can see from the definition of Perplexity that it essentially is the size of the set of words from which the next word is chosen given that we observe the history of words spoken. This is highly dependent on the domain of discourse, with formal texts usually having a lower value of perplexity. As can also be seen from the definition, the Perplexity value is improved if the Language Model assigns greater probability mass to words which are actually used in correct order. So, if we assume that we have a good enough Language Model, we can say that a sentence has good fluency if it has a lower value of perplexity. \\\\
\subsection{BLEU Score}
BLEU (Bilingual Evaluation Understudy) is an algorithm for evaluating the quality of text which has been machine-translated from one natural language to another. Quality is considered to be the correspondence between a machine's output and that of a human: "the closer a machine translation is to a professional human translation, the better it is". BLEU was one of the first metrics to achieve a high correlation with human judgments of quality, and remains one of the most popular.\\\\
Scores are calculated for individual translated segments - generally sentences - by comparing them with a set of good quality reference translations. Those scores are then averaged over the whole corpus to reach an estimate of the translation's overall quality. Intelligibility or grammatical correctness are not taken into account. BLEU is designed to approximate human judgment at a corpus level, and performs badly if used to evaluate the quality of individual sentences. BLEU’s output is always a number between 0 and 1. This value indicates how similar the candidate and reference texts are, with values closer to 1 representing more similar texts.\\\\
BLEU uses a modified form of precision to compare a candidate translation against multiple reference translations. The metric modifies simple precision since machine translation systems have been known to generate more words than appear in a reference text. This is illustrated in the following example~\footnote{Example Taken from Wikipedia}:\\\\
\textbf{Example of Machine Translation output with high precision}\\
Candidate: the the the the the the the\\
Reference 1: the cat is on the mat\\
Reference 2: there is a cat on the mat\\
Of the seven words in the candidate translation, all of them appear in the reference translations. Thus the candidate text is given a unigram precision of:\\
$P = \frac{m}{w_{t}} = \frac{7}{7} = 1 $\\\\
where $m$ is number of words from the candidate that are found in the reference, and $w_{t}$ is the total number of words in the candidate. This is a perfect score, despite the fact that the candidate translation above retains little of the content of either of the references.\\\\
The modification that BLEU makes is fairly straightforward. For each word in the candidate translation, the algorithm takes its maximum total count, $m_{max}$, in any of the reference translations. In the example above, the word "the" appears twice in reference 1, and once in reference 2. Thus $m_{max} = 2$.\\\\
For the candidate translation, the count $m_w$ of each word is clipped to a maximum of $m_{max}$ for that word. In this case, "the" has $m_{w} = 7$ and $m_{max}=2$, thus $m_{w}$ is clipped to 2. $m_{w}$ is then summed over all words in the candidate. This sum is then divided by the total number of words in the candidate translation. In the above example, the modified unigram precision score would be:\\
$P = \frac{2}{7} $\\\\
The above method is used to calculate scores for a range of n-gram lengths. The length which has the "highest correlation with monolingual human judgments" was found to be four. The unigram scores are found to account for the adequacy of the translation, how much information is retained. The longer n-gram scores account for the fluency of the translation, or to what extent it reads like "good English". \\\\
The modification made to precision does not solve the problem of short translations, which can produce very high precision scores, even using modified precision. An example of a candidate translation for the same references as above might be:\\
the cat\\\\
In this example, the modified unigram precision would be,\\
$P = \frac{1}{2} + \frac{1}{2} = \frac{2}{2}$\\\\
as the word 'the' and the word 'cat' appear once each in the candidate, and the total number of words is two. The modified bigram precision would be $\frac{1}{1}$ as the bigram, "the cat" appears once in the candidate. It has been pointed out that precision is usually twinned with recall to overcome this problem , as the unigram recall of this example would be $\frac{2}{6}$ or $\frac{2}{7}$. The problem being that as there are multiple reference translations, a bad translation could easily have an inflated recall, such as a translation which consisted of all the words in each of the references.\\\\
In order to produce a score for the whole corpus the modified precision scores for the segments are combined, using the geometric mean multiplied by a brevity penalty to prevent very short candidates from receiving too high a score. Let $r$ be the total length of the reference corpus, and $c$ the total length of the translation corpus. If $c \leq r$, the brevity penalty applies, defined to be $e^{(1-\frac{r}{c})}$. (In the case of multiple reference sentences, $r$ is taken to be the sum of the lengths of the sentences whose lengths are closest to the lengths of the candidate sentences. 
\section{System Overview}
The entire experimental procedure can be divided into the following components:
\begin{enumerate}
\item{Obtaining noisy natural language sentences.}
\item{Constructing the language model.}
\item{Obtaining the best substitution for an input noisy sentence.}
\end{enumerate}
Out of the number of sources mentioned previously from which noisy data might be made available, we have used poor machine translation for our experiments. In order to obtain the noisy data, we used a Statistical Machine Translation (SMT) system trained on a very small amount of data. The required language model for the translation system was derived using this small amount of data. This data was fed to GIZA++~\cite{GIZA} to learn the translation model. The language model was created using the SRILM language modeling toolkit~\cite{SRILM}. The translation model and the language model were then fed to MOSES~\cite{moses} decoder which translated the candidate source language sentences into target language sentences. These translated sentences are the noisy sentences that we attempt to correct in the next step.\\\\
Next, we use a large target language monolingual corpus of the in order to construct the language model for correction. Once again, SRILM is used to compute the language model in ARPA language model format. This language model along with the noisy sentences are then passed on to the fluency improvement algorithm in order to correct the noisy sentences.


\chapter {Fixed Length Phrase-based Correction Model}
Finding an optimal phrase segmentation is the most integral part of our task. In the first approach, we try to use fixed length phrases for this purpose. The replacement mechanism we follow is local to the phrases themselves, and the replacements made to a phrase do not in any way affect the replacements made to any other phrase. 
\section{Algorithm}
The first approach is the Fixed Length Phrase-based approach. In this approach, we divide the sentence into a set of phrases of fixed length and compute the best substitute for each phrase using overlapping n-grams. The algorithm then performs a set of local replacements on each phrase. Let us consider an example to see how this algorithm works:\\\\
Let the phrase be $P=p_1 p_2 \ldots p_7$\\
Let the language model have order = 4\\
Initially we have, starting index equal to 1. So first, consider the subphrase $p_1 p_2 p_3 p_4$.\\
Let the set of replacements for this phrase be:
\begin{enumerate}
\item $r_1 = x_1 x_2 x_3$
\item $r_2 = y_1 y_2 y_3 y_4$
\end{enumerate}
First, replace $p_1\ldots p_4$ with $r_1$. We obtain $P'=x_1 x_2 x_3 p_5 p_6 p_7$.\\
Next, make a recursive call to the function with $P'$ as the phrase to be replaced and start index equal to 2. Carry out these recursive calls until the entire phrase $P$ is replaced and compute the score. Next, perform a backtracking up the recursion stack and replace with the other available choices. When the control comes back to the top level function call, replace $p_1\ldots p_4$ with $r_2$ and follow the same recursive algorithm. The sequence of recursions that give the best score for the replaced phrase is taken to be the replacement for the phrase $P$. This process is repeated for all phrases.\\\\
\begin{algorithm}
\begin{algorithmic}
\REQUIRE Language model $LM$
\REQUIRE $n=$ Order of $LM$
\REQUIRE $L=$Size of phrases
\REQUIRE Phrase $P=w_1w_2\ldots w_L$
\REQUIRE global $current\_best=Score_{LM}(P)$
\REQUIRE global  $best\_sub=P$
\STATE recursive  function {\tt COMPUTE\_SUB}$(P, start\_index, Score):$
\IF {$start\_index+n>L$}  \STATE \COMMENT {The end of phrase has been reached}
\IF {$LMScore>current\_best$}
\STATE $current\_best=Score$
\STATE $best\_sub=P$
\ENDIF
\ELSE
\STATE {$kbest=${\tt FIND\_K\_BEST}($w_{start\_index},\ldots w_{start\_index+n-1}$)}
\FORALL {$x \in kbest$}
\STATE $P'=$Substitute $w_{start\_index}\ldots w_{start\_index+n-1}$ with $x$
\STATE {\tt COMPUTE\_SUB}$(P' ,start\_index+1, Score+Score_{LM}(P'))$
\ENDFOR
\ENDIF
\end{algorithmic}
\caption{Algorithm involving fixed length phrases to compute best substitution}
\label{algo:Fixed}
\end{algorithm}
In order to find the best replacement for a phrase, we use a very simple heuristic. We choose a phrase $Q$ to be a possible replacement for a phrase $P$ if the two phrases share at least 2 common words. The algorithm is formalized in Algorithm~\ref{algo:Fixed}. The function {\tt FIND\_K\_BEST()} takes in a phrase as argument and returns all phrases in the language model which has at least 2 words in common with the given phrase. The algorithm uses a global variable called $cur\_best$ which stores the best substitute recorded till that instant. The steps followed for each phrase $P$ are as follows:
\begin{enumerate}
\item Initialize $cur\_best$ as the phrase $P$ and $Score$ as the score of $P$ according to language model.
\item Call the function {\tt COMPUTE\_SUB}($P$,0,$Score$).
\item Store the string returned in step 2 as the replacement for the phrase $P$.
\end{enumerate}
After all the phrases have been substituted, the algorithm reconstructs the final sentence by simply concatenating the phrase substitutions. After it does so, it computes the score of the entire sentence according to the language model. If the score is greater than the score of the original sentence, it outputs the resulting sentence, otherwise outputs the original sentence.\\\\
As can be seen, this algorithm works locally on phrases and does not consider the relation between two phrases based on their location in the candidate sentence. However, within phrases, it follows a continuous substitution policy to determine the best substitution. The substitutions within one phrase are overlapping. An important feature of this algorithm is the fact that it does not bring about any loss in the quality of sentences. Due to the fact that we compare the best substitution with the original sentence, the resulting sentences are at least as good as the original ones. 

\section{Experiments and Results}
For our experiments, we used English as the target language. As previously mentioned, due to lack of data from other domain, we are using data from poor machine translation systems in order to check our algorithms. For testing the performance of this approach, we used the Europarl~\cite{EUROPARL} Corpus. We used 1000 sentences from the Europarl English-German bitext in order to train the machine translation system. 100 English sentences comprised the test set. In order to build the language model, we used the entire English part of the Europarl corpus. For perplexity computation, we used the perplexity computation module of SRILM. \\\\
For experimentation, we used a 4-gram Language Model, i.e., $n=4$. The length of phrase size was chosen to be 7. The results obtained from evaluating over the 100 sentences in the test set are summarized in table \ref{Table:overlap}.\\\\
We can see that this algorithm does not do well. In fact, we see no change in perplexity or BLEU score, signifying that it has, in fact, not made any change to the set of sentences. Therefore, this algorithm does not bring about any change to the quality of sentences in the test set. An analysis of the sentences used as the test set reveals the same fact. The output sentences were exactly the same as the input sentences.\\\\
\begin{table}
\begin{tabular}{|l|l|l|}
\hline
 & Perplexity & BLEU\\
\hline
Before Correction & 20.419 & 0.662\\
\hline
After Correction & 20.419 & 0.662\\
\hline
\end{tabular}
\caption{Results of the fixed length phrase-based correction algorithm}
\label{Table:overlap}
\end{table}
\section{Analysis}
An analysis of the steps of this algorithm reveals the fact that this algorithm probably performs too many substitutions, thereby actually attempting to increase the amount of noise in the sentences. Also, human beings do not usually attempt to make changes in sentences in this fashion. Corrections are made on phrases. One phrase is taken as a single unit, and some changes are made in that phrase so that the resultant phrase resembles the actual phrase in some way, but the errors of the original phrase are eliminated. Also, this algorithm fails to deal with inter-phrasal correlation, for which a conditional probability among phrases should have been used. All these factors contribute to the failure of the algorithm.\\\\
Another factor that does not support the use of this algorithm is the time complexity. for each phrase, the algorithm takes time which is exponential in the number of replacement candidates. In order to understand the time complexity of the algorithm, let us model the phrase replacement procedure as a graph. Every possible phrase is a node and there exists an edge from one node $v_i$ to another node $v_j$ if there exists a possible substitution which can transform $v_i$ into $v_j$. This algorithm tries to explore all possible sequence of substitutions to transform a sentence. Hence, in the graph model, it tries to explore all possible paths from a starting node. Since a graph can have exponential number of such paths, this algorithm has an exponential time complexity which renders it useless for almost all practical purposes.


\chapter{The Dynamic Programming based method}
\section{Motivation}
The failure of the fixed length phrase-based approach led us to conclude that this approach was only trying to increase the amount of error already present in the sentences by trying to make too many substitutions. So we hypothesize that an approach that takes in a phrase as one unit and makes corrects a phrase in one step after identifying the errors might be a good way to proceed. Also, the previous approach did not have any function that explicitly models all the previously mentioned errors. As a result, in this approach, we also modify the function used to select the replacements fro the phrases. Another problem with the previous approach is its exponential time complexity and the fact that it consumes much memory due to its recursive nature. We need to find a more efficient way of selecting the best phrase segmentation of a sentence. We also want to allow variable length phrases. All this can be accomplished by a simple recursive formulation of the problem as follows:\\
Suppose, we have a sentence $ S = s_1 s_2 \ldots s_n$\\
Suppose, we wish to find the best replacements for the phrase $S_{ij} = s_i\ldots s_j$\\
Suppose, $|$ denotes the concatenation operator. Then, we can write, $S_{ij} = S_{il}|S_{(l+1)j}$ for any $i\leq l\leq j$. \\
Hence, the task of computing $S_{ij}$ reduces to computing the best $l$ in accordance with the above equation. Here, we define "best" as the $l$ for which the replacements have highest score according to language model. Iteratively carrying out this operation would give us the replacements for $S_{1n}=S$.\\\\
We can easily see that this can be modeled as a bottom up dynamic programming problem. If we have replacement candidates for $S_{il}$ and $S_{(l+1)j}$, we can combine them using the above equation in order to get $S_{ij}$. This formulation is, in fact, identical to the formulation of the famed Cocke-Younger-Kasami (CYK) algorithm~\cite{cyk1}\cite{cyk2}\cite{cyk3} for parsing of context-free grammars. Once we have the initial set of replacements for each $S_{ij}$ computed using the distance function mentioned subsequently, we can use this formulation to combine them all in order to find the best substitute for the entire sentence. An important point to note with variable phrase lengths is the fact that we need to explore all possible phrasal decompositions, whose number is exponential in the length of the sentence. However, this formulation allows us to compute the best decomposition in time polynomial in the length of the sentence. Thus, this formulation gives us a more efficient way of solving the problem.
\section{Algorithm}
Suppose we have a sentence $S$ which we wish to correct. Say $S$ has $N$ words $S_1,\ldots,S_N$. So, \\
$S=S_1 S_2 \ldots S_N$\\\\
This sentence may be divided into an $O(N^2)$ number of contiguous phrases. Each phrase may begin at any $S_i$ and end at any $S_j$ satisfying the constraint $1\leq i\leq j\leq N$. We denote a phrase beginning at $S_i$ and ending at $S_j$ as $P_{ij}$. Let the Language Model be denoted by $LM$. The algorithm initially computes a list of substitutes for each $P_{ij}$ from $LM$. For this purpose, a function {\tt FIND\_BEST\_SUB()} is called. It takes in a phrase $P$ and returns a list of $k$ best replacement phrases for $P$ taken from $LM$. For now, we treat this function as a black box. This function is discussed in detail in the next section. \\\\
Let the list of top-$k$ substitutes for $P_{ij}$ be denoted by $SUB_{ij}$. $SUB_{ij}$ constitutes the initial list of substitutes taken from the Language Model. Each entry of $SUB_{ij}$ is a 2-tuple of the form (phrase, score) where the score of the phrase is the logprob value according to $LM$. We combine these lists using a Dynamic Programming algorithm. We fill a table $REP$. It is a $N\times N$ table. The entry $REP_{ij}$ consists of the top $k$ substitutes for $P_{ij}$ taken from $SUB_{ij}$ and other entries of the table $SUB$. $REP_{ij}$ is computed as follows:\\
\begin{enumerate}
\item $REP_{ij}=SUB_{ij}$
\item $\forall l, i\leq l \leq j-1,$ we have $k^2$ candidate substitutes for $P_{ij}$ because there are $k$ candidates present in each of $SUB_{il}$ and $SUB_{(l+1)j}$. For each of these $k^2$ candidates, Compute the score of the phrase $s_{ii}|s_{(l+1)j}$ according to $LM$, where $s_{ik}\in SUB_{ik}$ and $s_{(k+1)j}\in SUB_{(k+1)j}$. Here $|$ denotes the string concatenation operation.
\item For each of the $k^2$ phrases constructed in Step 2, add each (phrase, score) pair to $REP_{ij}$.
\item Retain the top $k$ entries in $REP_{ij}$ based on the scores.
\end{enumerate}
\begin{algorithm}
\begin{algorithmic}
\STATE sentence $S =S_1 S_2\ldots S_N$
\STATE phrase $P_{ij}=S_i\ldots S_j.$ We have $S=P_{1N}$
\STATE $SUB=N\times N$ array storing the initial replacements. $SUB_{ij}$ has replacements for $P_{ij}$
\STATE $REP=N\times N$ array storing the final replacements. $REP_{ij}$ has replacements for $P_{ij}$
\STATE $\forall i,j 1\leq i\leq j\leq N$, compute $SUB_{ij}=$ {\tt FIND\_BEST\_SUB}($P_{ij}$)
\FOR {$i = 1$ to $N$}
\STATE {$REP_{ii}=SUB_{ii}$   //Fill up the table for all strings of length one}
\ENDFOR
\STATE{//Length of substrings being replaced}
\FOR {$l=2$ to $N$ }
\STATE{//Starting position of the string being replaced}
\FOR{$i=1$ to $N-l+1$}
\STATE{//Iterate over all possible break points}
\FOR {$j=i$ to $i+l-1$ }
\STATE{//Initialize $REP_{i(i+l-1)}$ to already computed replacements}
\STATE {$REP_{i(i+l-1)} = SUB_{i(i+l-1)}$}
\FORALL {(p,s) $\in SUB_{ij}$}
\FORALL {(p',s') $\in SUB_{(j+1)(i+l-1)}$}
\STATE {$q=p|p'$, where $|$ is the concatenation operation}
\STATE {$r$=score($q$)}
\STATE {$REP_{i(i+l-1)}=REP_{i(i+l-1)}\cup (q,r)$}
\ENDFOR
\ENDFOR
\ENDFOR
\STATE {$REP_{i(i+l-1)}=$ Top $k~(q,r)$ pairs having best scores from $REP_{i(i+l-1)}$}
\ENDFOR
\ENDFOR
\end{algorithmic}
\caption{Dynamic Programming based correction algorithm}
\label {DisjPhr}
\end{algorithm}
\begin{figure}
\includegraphics[width= 80mm, height = 70mm] {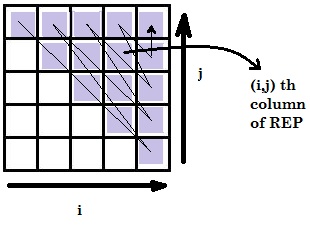}
\label{fig:DP}
\caption {An illustration of how the Dynamic Programming algorithm proceeds for $N=5$}
\end{figure}
The entries of $REP_{1N}$ give us the final optimum substitution for the sentence $S$. The algorithm is formally stated in Algorithm \ref{DisjPhr}.\\\\
The algorithm fills up a $N\times N$ table. First, it fills up the entries corresponding to phrases of length 1, then phrases of length 2 and so on. Let us consider a small example to see how the algorithm works:\\\\
Say, we are trying to fill up $REP_{ij}$. Initialize $REP_{ij}=SUB_{ij}$.\\
Say, we take a division at some $l$.\\
Let, $REP_{il}=\{P_1,P_2,\ldots, P_k\}$ and $REP_{(l+1)j}=\{Q_1,Q_2,\ldots ,Q_k\}$.\\
The strings we add to $REP_{ij}$ are $\{P_1Q_1, P_1Q_2,\ldots ,P_1Q_k, P_2Q_1,\ldots ,P_2Q_k,\ldots ,P_kQ_1,\ldots ,P_kQ_k\}$. Then, we retain the top $k$ best strings from the set $REP_{ij}$ according to the scores assigned to the strings by the language model.\\
Figure~\ref{fig:DP} gives us an illustration of the way the algorithm works. The shaded area of the table is the one which is actually filled during the course of the algorithm. The arrows depict the direction in which the table is filled.\\\\
The algorithm runs in polynomial time once the function {\tt FIND\_BEST\_SUB()} has been computed. The Complexity of the remaining part of the algorithm stems from the 5 nested \textbf{for} loops. The three outermost loops run in $O(N)$ time each and the two innermost loops run in $O(k)$ time each, giving this loop a time complexity of $O(N^3k^2)$. So, the overall time complexity of this algorithm is $O(N^3k^2)$. In addition, the function {\tt FIND\_BEST\_SUB()} is called for $O(N^2)$ times.\\\\
A naive implementation of the function {\tt FIND\_BEST\_SUB()} would be to perform a complete linear scan of the entire list, match the target phrase with each phrase in the list and select the best matches. This would be very inefficient, because of the extremely large number of phrases in the language model. Hence, we need a more efficient method of implementing this function. This implementation issue has been handled in the next section.
\section{Computing the Phrase Substitutes}
The phrasal substitutes have been computed using the {\tt FIND\_BEST\_SUB()} function. This function takes in a single argument (the phrase $p$) and returns a list of 2-tuples of the form ($q$,$s$) where $q$ is a candidate replacement for $p$ and $q$ has a score of $s$ according to the language model being used. As mentioned in the previous section, a naive implementation of this function using linear search and a distance function would result in very high time complexity. So, an efficient implementation is essential. \\\\
This problem can be modeled as an Information Retrieval task. We can treat each phrase in the language model as a document and the replacement candidate as the query. Our aim is to find out which documents best match the query and also have high scores. In order to compute the $k$-best list, we follow a 2-step process:\\
\begin{enumerate}
\item From all the phrases in the language model, compute the top $T$ best phrases that match the replacement candidate.
\item From the $T$ phrases computed in the previous step, compute the $k$ phrases having the best scores.
\end{enumerate}
As in every Information Retrieval system, we need to first index the documents. Each phrase in the language model is treated as an individual document and assigned a {\tt docid}. Next, an inverted index is constructed based on the words present in the phrases. The inverted index is similar to the index used in standard Information Retrieval systems. Corresponding to every word, there is a postings list of {\tt docid} in which that word occurs. A dictionary of words is also maintained. In order to facilitate implementation, the words in the dictionary are stored in  a prefix-tree (trie), which significantly lowers the search complexity for the dictionary. \\\\
During the retrieval phase, we use a distance function between words in order to match {\tt docid} to the query. Let this distance function be $D(w_1,w_2)$. The retrieval process can be formally expressed as follows:\\\\
Say, the query $q$ contains $w_q$ words $q_1, q_2, \ldots ,q_{w_q}$. \\\\
Let us say, by applying $D(q_i,w_j)$ for all $w_j$ in the dictionary, we get the set $S_q={s_1,s_2,\ldots ,s_x}$ to be the set of words that are similar to the words in $w_q$. \\\\
Let the postings list for each $s_i$ in $S_q$ be represented by $l_i$. \\\\
Then the list of documents which satisfy the query is given by $Out_q=\cup_{i=1}^x l_i$. Thus, $Out_q$ is the list of phrases that have words which are in some way similar to the query replacement candidate phrase $q$. \\\\
In order to carry out the intersection in the previous step efficiently, the {\tt docid}s in each postings list is maintained in sorted order. Such representations in the form of a linked list give us an union time which is linear in the size of the two lists. \\\\
Once we get the list of phrases which have words similar to the words in the replacement candidate phrase, we can do a simple linear scan to find out the best $T$ phrases from this list. This is because, in practice, we have $|Out_q| << M$. We use a phrase level distance function computing the $T$-best list. A number of facts were taken into consideration while defining this distance function. The distance function is a linear combination of different components catering to different possible sources of error. They are:
\begin{enumerate}
\item Orthographic Errors, like typographical errors etc.
\item Word errors, for example, using words out of context or at unsuitable positions.
\item Reordering based errors, most commonly seen in the case of Machine Translation from Free Word Order languages to Fixed Word Order languages.
\end{enumerate}
The different components of the distance function are as follows:
\begin{enumerate}
\item {Levenshtein distance~\cite{levenshtein} between individual words ($f_1$), which models the orthographic errors.}
\item {Synset Distance in Wordnet ($f_2$), which models the incorrect use of synonyms in sentences.}
\item {Word order based metrics ($f_3$), which models the errors created by different word order in different languages.}
\end{enumerate}
Distance function ($f_d$) can be represented as a linear combination of the above metrics:\\
$f_d = \sum_{i=1}^{4} \alpha_if_i$\\\\
Each of these metrics is explained in detail in the following subsections.
\subsection{Levenshtein distance between individual words}
Levenshtein Distance is a commonly used metric in tools like spell checkers. It is commonly known as the Edit Distance algorithm. It is used to find out the distance between two words. Levenshtein distance measures the minimum number of changes required in order to transform one word into another. For this, it permits the following operations:\\
\begin{enumerate}
\item {\textbf{Insertion}: It refers to inserting a new character to transform the word. For example, cat$\rightarrow$ca\textbf{r}t. In this, \textbf{r} is inserted between \textbf{a} and \textbf{t} in the original word.}
\item {\textbf{Deletion}: This operation is essentially the reverse of insertion, where we delete an already present character in  the original word to form a new word. For example, ca\textbf{r}t$\rightarrow$cat. In this, the \textbf{r} between \textbf{a} and \textbf{t} is deleted in the original word.}
\item {\textbf{Substitution}: This operation involves the replacement of one character in the original word with another character. For example, ca\textbf{r}t$\rightarrow$ca\textbf{s}t. Here, the \textbf{r} in the source word is replaced by \textbf{s}. This operation can be seen as a combination of insertion and deletion. The previous example can be expressed as an insertion followed by a deletion, like, cart$\rightarrow$carst$\rightarrow$cast, or as a deletion followed by an insertion, like cart$\rightarrow$cat$\rightarrow$cast. However, we use this as a separate metric in order to lower the edit distance value. }
\end{enumerate}
The Levenshtein distance measure is used in the algorithm at two stages:\\
\begin{enumerate}
\item {Firstly, this metric is used in order to fetch the phrases which seem relevant to the replacement candidate. This is used as a part of the distance function $D(w_1,w_2)$. Say, for some $q_i\in w_q$, and some $w_j$ in the dictionary, we have $D(q_i,w_j)<D_t$ ($D_t$ is a predetermined threshold). Then we include $w_j$ in $S_q$. }
\item {Secondly, this metric is used in the computation of the $k$-best list of replacements as part of functions $f_1$ and $f_4$. The details are as follows:\\
\begin{enumerate}
\item {\textbf{As part of $f_1$}:\\ Say, we are given a phrase $P=p_1,p_2,\ldots p_n$ and a possible replacement for that phrase $R=r_1,r_2,\ldots r_m$. We compute the Levenshtein distance between each $(p_i,r_j)$ pair. We normalize this value by the maximum of the lengths of $p_i$ and $r_j$.\\Now we define the score for the phrase $R$ as the minimum normalized edit distance for a word in $P$ with words of $R$ summed over all the words in $P$. Formally speaking,\\\\ $f_1 = \sum_{p\in P} max_{r\in R}\frac{LevenshteinDistance(p,r)}{max(len(p_i),len(r_j))}$}
\item {\textbf{As part of $f_4$}:\\ In the function $f_3$, the Levenshtein distance metric is used in order to preprocess the phrases before computing the actual word order based distance metrics. The preprocessing phase actually performs an alignment between the words in the replacement candidate and the target phrase. As in the previous case, let us consider P to be the phrase to be replaced and R to be a candidate replacement phrase. A word $p \in P$ is said to be aligned to a word $r \in R$ iff the Levenshtein distance between $p$ and $r$ is less than some threshold. In this phase, we perform a one-to-one alignment, with ties broken arbitrarily. That is, if two words in $P$ get aligned to a single word in $R$, we choose only one of the words of $P$ to be aligned and ignore the other one.}
\end{enumerate}}
\end{enumerate}
\subsection{Synset Distance in Wordnet}
Almost all languages have an extremely rich vocabulary. As a result, it becomes almost impossible for a single person to remember all words in a language. So, it is quite often the case that words are used in an incorrect fashion. For example, quite often it is the case that a word is used in a sentence while one of it's synonyms or a word closely related to it has to be used in its stead. Let us consider the following sentence:\\\\
\textbf{He was a one game surprise.}\\\\
However, a more correct and semantically apt sentence would be:\\\\
\textbf{He was a one game wonder.}\\\\
Such inaccuracies can be taken care of only when the words in question are replaced by more suitable words. That is where the synset based word matching part of the distance function helps us compute better replacements for a word in a phrase.\\\\
In order to compute this metric, we use the Wordnet~\cite{wordnet}. WordNet is a lexical database for the English language. It groups English words into sets of synonyms called synsets, provides short, general definitions, and records the various semantic relations between these synonym sets. The purpose is twofold: to produce a combination of dictionary and thesaurus that is more intuitively usable, and to support automatic text analysis and artificial intelligence applications. The database and software tools have been released under a BSD style license and can be downloaded and used freely. The database can also be browsed online.\\\\
Each synset in the wordnet consists of words having similar semantics, or in short, synonyms. The wordnet also supports a number of other relationships between synsets. In fact, we can view it as a graph where the vertices are the synsets and the edges are the relationships among the synsets. Some of the relationships supported by the Wordnet are:\\
\begin{enumerate}
\item Hyponymy
\item Hypernymy
\item Meronymy
\item Holonymy
\end{enumerate}
In this work, only the synonyms have been used. As before, let us consider the replacement candidate to be $P$ and a possible replacement to be $R$. For each word in $P$, we check if there is any word in $R$ which belongs to the same synset. We define a Boolean function $g(p,R)$ which takes the value of $1$ if there is any $r \in R$ belonging to the same synset as $p$ in the Wordnet, and $0$ otherwise. The function $f_2$ is defined as the sum of the function $g(p,R)$ for all words $p\in P$ and normalized by the length of $P$. Therefore, we have,\\\\
$f_2 = \frac{\sum_{p\in P}g(p,R)}{len(P)}$
\subsection{Word order based metrics}
Word order becomes an important factor, especially in translation tasks. For example, in a language like Bengali or Hindi, the order of words is Subject-Object-Verb, whereas in English it is Subject-Verb-Object. So, a common problem in translation systems is the lack of a good reordering model. Also, when non-native speakers or new learners of a language try to write sentences in the language, word order errors are some of the more commonly observed errors. Hence, this is a problem worthy of consideration. And that is why this factor in the distance function is extremely important. If this factor is not kept, phrases might be chosen which have an incorrect word order but higher overall score. This factor takes care of the fact that correct word order is preserved when selecting replacement phrases. \\\\
In this work, two types of functions have been used for word order preservation:\\
\begin{enumerate}
\item {\textbf{Rigid Word Order}: As mentioned in the subsection on Levenshtein distance, first a word level alignment is performed between the phrases $P$ and $R$. Next, it is checked if the words in $R$ aligned to $P$ are in the same order as the corresponding words in $P$. The phrase $R$ is considered for replacement if and only if the aligned words are in the same order in both $P$ and $R$. Otherwise, the phrase $R$ is discarded.}
\item {\textbf{Flexible Word Order}: The rigid word order metric does not always work. In fact, a certain degree of flexibility is required so that word order related problems can be taken care of. But it must be ensured that the flexibility is not absolute, otherwise, it might lead to the introduction of more noise in the sentences. In this case too, we consider the previous model where a phrase $R$ is a possible replacement candidate for a phrase $P$. Using the word based Levenshtein distance algorithm, first an alignment is computed. Each word pair in the phrases is tagged. Next, either one of these two metrics is computed:\\
\begin{enumerate}
\item Length of Longest Common Subsequence: The length of the Longest Common Subsequence (LCS) is computed between the aligned words in $R$ and $P$. This value is normalized by the number of aligned words and that gives us the value of $f_3$. 
\item Number of Inversion Pairs: Given an array $A=[A_1,A_2,\ldots ,A_n]$, $A[i]$ and $A[j]$ is said to be an inversion pair iff $i>j$ but $A[i]<A[j]$. In this method too, the words are aligned and the number of inversion pairs in $R$ are computed. The inverse of the number of inversion pairs gives the value of $f_3$.
\end{enumerate}
Let us consider an example:\\
Say, $ R = r_1,r_2,r_3,r_4$\\
And,$ P = p_1,p_2,p_3,p_4,p_5$\\
Say, after the alignment phase, we have $p_1$ aligns to $r_2$, $p_2$ aligns to $r_4$ and $p_4$ aligns to $r_3$.\\
So, the number of aligned pairs is 3.\\
Let us tag the pairs with numbers and ignore the rest of the words for the time being. So, the phrases become:\\
$P = 1,2,3$\\
$R = 1,3,2$\\
If we consider the LCS metric, the value will be $f_3 = \frac{2}{3}$, as {2,3} is a possible Longest Common Subsequence of $P$ and $R$. \\
If we consider the Inversion metric, the value will be $f_3 = \frac{1}{3}$. {3,2} is an inversion pair in $R$.}
\end{enumerate}
\section{A Representative Example}
Let us consider the following sentence:\\\\
\textbf{the europe extreme right in is characterized by its and its use of immigration as differences the issue of.}\\\\
We can see that there are a number of errors in this sentence. For example:\\
\begin{enumerate}
\item The phrase \emph{europe extreme right} should become \emph{european extreme right} or \emph{extreme right of europe}.
\item The word \emph{in} should not be there after \emph{right}. 
\item The phrase \emph{its and its use}. There should either be a word after the first \emph{its}, or the phrase should be \emph{its use}.
\item The phrase \emph{differences the issue of} makes no sense. This phrase should be reordered or replaced by something else.
\end{enumerate}
Let us now see how the algorithm works on this sentence. \\\\
Firstly, it calls on the function {\tt FIND\_BEST\_SUB()} in order to compute the best replacements for all possible phrases. We have used the 5 best replacements in this experiment. A snapshot of the replacement set of some of the more significant phrases is given below along with the scores. Each replacement is written as a (phrase,score) pair. \\\\
\textbf{$SUB($europe$)$} = \{(europe, 5.9865), (european, 5.9012), (european union, 4.5213), (the european life, 2.3242), (european governments, 2.2312)\}\\
\textbf{$SUB($extreme$)$} = \{(extreme, 1.2121), (extremist, 0.9821), (extremist wings, 0.5632), (the islamic extremists, 0.2180), (fundamentalists and extremists, 0.1134)\}\\
\textbf{$SUB($europe extreme$)$} = \{(european, 5.1212), (european extreme, 4.9212), (extremist threat, 4.2911),(threat to europe, 3.2931), (terrorism in europe, 3.1023)\}\\
\textbf{$SUB($europe extreme right$)$} = \{(european extreme right, 4.2313), (extreme europe right, 4.0981),(european right wing, 3.5673),(european liberal politics, 3.2342),(the european union, 3.2312)\}\\
\textbf{$SUB($extreme right in$)$} = \{(extreme right is, 3.4562), (extreme right, 3.2576), (rightist extremists are, 2.7853), (in the extremist propaganda, 1.0123), (right to say, 0.4323)\}\\
\textbf{$SUB($its and its use$)$} = \{(its use, 5.2142), (that it uses ,4,2123), (its diplomacy and its, 4.0123), (that it had used, 3.1463), (they had been used, 2.6432)\} \\
\textbf{$SUB($as differences the$)$} = \{(make good the differences, 2.1231),(as the differences, 1.4654), (differences of the, 1.2435), (different from the usual, 0.3461), (different issues in the, 0.2423)\}\\
\textbf{$SUB($differences the issue of $)$} = \{(issue of the differences, 2.3242), (address the issues of differences, 2.1982), (address the matter of, 1.5692), (the differences among the, 1.2342), (the problems of different, 0.8734)\}\\\\
The length of the sentence is $N = 19$. So, the algorithm creates a $19\times 19$ matrix and fills up its diagonal and the upper triangle. First it fills up the diagonal for all the unigrams in the sentence, then all the bigrams and so on. In order to see how this algorithm actually works, let us consider the phrase \textbf{the europe extreme right in}. This phrase can be broken down in the following ways:
\begin{enumerate}
\item the + europe extreme right in
\item the europe + extreme right in
\item the europe extreme + right in
\item the europe extreme right + in
\end{enumerate}
After the lower rows of the table is filled up, the cell corresponding to the phrase \textbf{the europe} and \textbf{extreme right in} looks as follows:\\
$REP($the europe$)$ = \{(the europe, 12.2341), (the european, 12.1914), (the the european union, 10.1231), (not in the european union, 9.1231), (the european governments, 6.2114)\}\\
$REP($extreme right in$)$ = \{(extreme right is, 3.4562), (extreme right, 3.2576), (rightist extremists are, 2.7853), (in the extremist propaganda, 1.0123), (right to say, 0.4323)\}\\\\
Combining these two, we get 25 candidates for replacing the phrase \textbf{the europe extreme right}, like:
\begin{enumerate}
\item \textbf{(the europe extreme right is, 16.0012)}
\item \textbf{(the europe extreme right, 15.8212)}
\item \textbf{(the europe right extremists are, 15.1231)}
\item (the europe in the extremist propaganda, 14.2111)
\item (the europe right to say, 12.9015)
\item \textbf{(the european extreme right is, 16.2918)}
\item \textbf{(the european extreme right, 15.9210)}
\item (the european rightist extremists are, 15.0012)
\item (the european in the extremist propaganda, 13.5681)
\item (the european right to say, 13.0001)
\item (the the european union extreme right is, 13.6221)
\item (the the european union extreme right, 13.5512)
\item (the the european union rightist extremists are, 12.9899)
\item (the the european union in the extremist propaganda, 11.5101)
\item (the the european union right to say, 10. 7014)
\item (not in the european union extreme right is, 12.8213)
\item (not in the european union extreme right, 12.5612)
\item (not in the european union rightist extremists are, 12.5235)
\item (not in the european union in the extremist propaganda, 11.2322) 
\item (not in the european union right to say, 9.4534)
\item (the european governments extreme right is, 10.9219)
\item (the european governments extreme right, 10.6212)
\item (the european governments rightist extremists are, 9.0028)
\item (the european governments in the extreme propaganda, 7.5422) 
\item (the european governments right to say, 6.8921)
\end{enumerate}
The 5 entries marked in bold in the above list mark the best 5 selections for the candidate phrase \textbf{the europe extreme right}. The same process is used to fill up the table $REP$, the best candidate replacement chosen for the entire sentence turns out to be:\\\\
\textbf{The European extreme right is characterized by its use of immigration as the issue of differences.}\\\\
As we can see, a number of changes have been made in the original sentence, like:
\begin{enumerate}
\item \emph{europe extreme right} $\rightarrow$ \emph{european extreme right}
\item \emph{its and its use} $\rightarrow$ \emph{its use}
\item \emph{differences the issue of} $\rightarrow$ \emph{the issue of differences}
\end{enumerate}
Quite clearly, the final sentence is more correct than the original sentence.
\section{Experiments and Results}
The variable part of the correction system described previously was the distance function used in order to find out suitable replacement candidates for the phrases. We conducted a number of experiments using several combinations of these distance functions. The used combinations are tabulated in Table~\ref{DistFn}.\\\\
\begin{table}
\begin{tabular}{|p{80pt}|p{80pt}|p{80pt}|p{80pt}|}
\hline
Function Code & $f_1$ & $f_2$ & $f_3$ \\
\hline
\textbf{A} & Levenshtein Distance & Synset Distance & - \\
\hline
\textbf{B} & Levenshtein Distance & Synset Distance & Rigid Word Order \\
\hline
\textbf{C} & Levenshtein Distance & Synset Distance & LCS Length \\
\hline
\textbf{D} & Levenshtein Distance & Synset Distance & Number of Inversion Pairs \\
\hline
\end{tabular}
\caption {The Distance functions used for experimentation}
\label {DistFn}
\end{table}
Experiments were performed using the Europarl Corpus. The languages chosen were German and English. Since we did not have sufficient amount of noisy data at our disposal, so we used a poorly tuned Statistical Machine Translation system in order to obtain the incorrect sentences. The source language was German and the target language was English. The correction was done on the English sentences. In fact, to make the task more difficult, a number of errors were manually incorporated into the incorrect English sentences to further increase the amount of noise present in the sentences. \\\\
The system is the same as explained in Section 2.4. In the implementation of the distance functions, there is one parameter that needs to be fixed. It is in the implementation of the Levenshtein Distance function, where a value of $D_t=3$ has been used. This value ensures that the substitute word chosen is not too different, nor too similar to the word in question. We have also provided equal weight to all the components of the distance function, that is, $\forall 1\leq i\leq 4, \alpha_i=0.25$. Table~\ref{Res} sums up the values of Perplexity and BLEU score for the different choices of the distance function.
\begin{table}
\begin{tabular}{|l|p{60pt}|p{40pt}|p{40pt}|p{40pt}|p{40pt}|}
\hline
  & Original Sentences & \multicolumn{4}{|c|}{Corrected Sentences}\\
\hline
& & Function A & Function B & Function C & Function D\\
\hline
Perplexity & 20.419 & 19.064 & 18.716 & 18.656 & 18.658\\
\hline
BLEU & 0.662 & 0.662 & 0.663 & 0.665 & 0.664\\
\hline
\end{tabular}
\caption { Summary of Results over all the used distance functions}
\label {Res}
\end{table}
In order to analyze the performance of our algorithm on Machine Translation, we peformed another set of experiments. For this purpose, we did not use a separate Machine Translation system, but used the Google Translator\footnote{http://translate.google.com/} instead. The corpus chosen was the FIRE \footnote{FIRE (Forum for Information REtrieval) is a conference conducted by ISI Kolkata. http://www.isical.ac.in/$\sim$clia/}2008 ad-hoc retrieval corpus. First, we selected 100 sentences from the English corpus. We translated those into Hindi using the Google Translator system and translated back the Hindi sentences into English using Google Translator again. Since English-Hindi translation in Google translator does not give us very good results, so a lot of noise was incorporated into the sentence. Let us take a look at some of the sentences in order to understand the amount of noise introduced:
\begin{enumerate}
\item {\textbf{Original:}Specifically CII has appreciated the special initiatives for agriculture, gems and jewellery, handlooms, leather and footwear.\\\\\textbf{Final:}CII agriculture particularly gems and jewelry, handicraft, leather and shoes for the special initiative is appreciated.}
\item {\textbf{Original:}Indian Chamber of Commerce president Anup Singh said, The special package for agriculture and schemes like Vishesh Krishi Upaj Yojna will boost exports of fruits, vegetables and their value-added products.\\\\\textbf{Final:}Indian Chamber of Commerce president, said Anoop Singh, Vishesh agricultural planning and agricultural schemes like the special packages for fruits, vegetables and their value-added products will boost exports.}
\item {\textbf{Original:}GMG plans to fly thrice a week between the south-eastern port city of Chittagong and Calcutta.\\\\\textbf{Final:}GMG to fly three times a week southeastern port city Chittagong and Kolkata between plans.}
\item {\textbf{Original:}GMG hopes to be allowed to fly to other Indian cities connecting capital Dhaka with Delhi and Mumbai.\\\\\textbf{Final:}GMG to other Indian cities Mumbai and Delhi to Dhaka link is expected to be allowed to fly.}
\item {\textbf{Original:}The government recently signed a free trade pact with Thailand, which it hopes will ultimately cover all nations in the region.\\\\\textbf{Final:}The government recently Thailand, which it hopes will eventually cover all countries in the region with a free trade treaty signed.}
\end{enumerate}
As is obvious, the final sentences contain a lot of errors, the most common being global clause reordering. Unfortunately, our system is not built to handle this problem. We used our noise correction algorithm with the Distance Function C on these sentences. We used the entire FIRE 2008 ad-hoc English corpus for computing our language model. The results are summarized in Table~\ref{FIRE2way}.
\begin{table}
\begin{tabular}{|l|l|l|}
\hline
 & Perplexity & BLEU\\
\hline
Before Correction & 27.521 & 0.557\\
\hline
After Correction & 16.121 & 0.569\\
\hline
\end{tabular}
\caption {Results of experiments with twice translated sentences taken from the FIRE corpus}
\label {FIRE2way}
\end{table}
Let us examine some of the output sentences from this experiment:
\begin{enumerate}
\item {CII agriculture particularly gems and jewelry handicraft leather and shoes for the special initiative is appreciated. $\rightarrow$ CII in agricultural particularly gems and jewelry handicraft leather and shoes special initiative has been appreciated. }
\item {Indian Chamber of Commerce president said Anoop Singh Vishesh agricultural planning and agricultural schemes like the special packages for fruits vegetables and their value added products will boost exports. $\rightarrow$ Indian Chamber of Commerce president Anoop Singh said Vishesh agricultural planning and agricultural schemes like the special packages for fruits vegetables and their value added products will boost revenue from exports.}
\item {GMG to fly three times a week southeastern port city Chittagong and Kolkata between plans. $\rightarrow$ GMG to fly three times a week between southeastern port city Chittagong and Kolkata.}
\item {GMG to other Indian cities Mumbai and Delhi link is expected to be allowed to fly. $\rightarrow$ GMG to other Indian cities like Mumbai and Delhi link is expected to be permitted to fly. }
\item {The government recently Thailand, which it hopes will eventually cover all countries in the region with a free trade treaty signed. $\rightarrow$ The Thailand government now hopes will eventually cover all countries in the region with a signed free trade treaty.}
\end{enumerate}
As we can see from the above examples, the global reordering problem has not been solved. But the algorithm has indeed made local changes to clauses and phrases and the output sentences are indeed better than the input sentences in terms of fluency. The semantics have not always been preserved, but that is not what we have aimed to do anyway. 
\section{Analysis}
The Dynamic Programming based correction model succeeds in improving the fluency of sentences quite significantly. This algorithm is quite similar to the standard decoding algorithms, where given a sentence, one is supposed to find a sequence of phrasal segmentations and using the phrase table, find a set of substitutes which would maximize the a-posteriori probability of the sentence being generated. In this algorithm, since we do not have any explicit phrase table, we implement it by searching for nearly matching phrases from the Language Model itself. The distance function helps figure out what is a good match. This algorithm does not use a global distortion model like is done in most Statistical Machine Translation systems. The rationale behind this is the fact that normally errors in writing text do not involve long-distance reordering errors, rather most of the wrongly written words occur within a small window. Our test set was so designed that it contained sentences having all the types of errors mentioned in Section 2.2. Some types of errors were obtained right from the translation output, while some were introduced manually.\\\\
A quick glance at Table~\ref{Res} shows that the order of words in the chosen replacement phrases is extremely important, as it significantly lowers the average perplexity of the set of sentences. Although the Distance Function A lowers the perplexity of the test set of sentences, indicating that correction of fluency has indeed taken place, the lower perplexities in case of Distance Functions B,C and D imply that total freedom in case of word order may not be the way to go. In fact, this is quite intuitive from the fact that English is not a Free Word Order language.\\\\
A look at the BLEU scores shows that they have not significantly changed in order to draw a conclusion about the algorithm's ability to alter the faithfulness of the sentences. However, a comparison between BLEU scores and Perplexity value shows that the BLEU score also follows the same trend as the perplexity values. Also, the Distance Functions C and D have resulted in almost comparable values of perplexity and BLEU, showing that using either of them would give us a good enough result. \\\\
Now let us look at some sentences that were corrected by this algorithm and try to analyze the errors that each distance function corrects:
\begin{enumerate}
\item {The parties dominant of the center left and center right have faced prospect this in the policy of the ostrich applying.\\\textbf{Distance Function A:} The parties dominant of the center left and center right have faced this prospect in the applying principle of the ostrich.\\\textbf{Distance Function B:} The dominant parties of the center left and right have faced this prospect in the policy of the ostrich applying.\\\textbf{Distance function C:} The dominant parties of the center left and right have faced this prospect in the policy of the ostrich applying.}
\item {This is precsely the aim pursued largely by research teams in economics, sociology, psychology and pol science in the United States.\\\textbf{Distance Function A:} That precisely is the aim largely pursued by teams of research in economics, sociology, physiology and political science in the United States.\\\textbf{Distance Function B:} This is precisely the aim pursued largely by teams of research in economics, sociology, psychology and political science in the United States. \\\textbf{Distance Function C:} This is precisely the aim pursued largely by teams of research in economics, sociology, psychology and political science in the United States.}
\item {This does not mean that it should be eliminated heterogeneity and create racially homogeneous communities.\\\textbf{Distance Function A:} This does not mean that it should eliminate heterogeneity and create homogeneous communities racially.\\\textbf{Distance Function B:}This does not mean that should eliminate heterogeneity and create racially homogeneous communities.\\\textbf{Distance Function C:}This does not mean that should eliminate heterogeneity and create racially homogeneous communities.}
\item {Other protest against action affirmative and maintains that a policy that about race does not care  is sufficient.\\\textbf{Distance Function A:}others protest against an affirmative action and maintains that a policy that about race does care is sufficient.\\\textbf{Distance Function B:}Others protest against affirmative action and maintains that a policy that does not about race care is sufficient.\\\textbf{Distance Function C:}Others protest against affirmative action and maintains that a policy that does not about race care is sufficient.}
\item {Obviously, minorities have made progressed towards more integration and economic success.\\\textbf{Distance Function A:}Obviously minorities have made progress towards more economic success and integration. \\\textbf{Distance Function B:}Obviously, minorities have made progress towards more integration and economic success.\\\textbf{Distance Function C:}Obviously, minorities have made progress towards more integration and economic success.}
\item {It will not be the case, as is clear the racial history of America.\\\textbf{Distance Function A:}It will not be the case, as is clear in the racial history of America.\\\textbf{Distance Function C:}It will not be the case, as is clear in the racial history of America.\\\textbf{Distance Function C:}It will not be the case, as is clear in the racial history of America.}
\end{enumerate}
As is apparent from the above examples, the Distance Functions B and C are almost identical in nature, and the changes performed by these functions is basically a superset of the changes performed by the Distance Function A. Distance Function A is able to correct substitution based errors, spelling errors and missing word errors, but may fail to correct reordering based errors. In fact, due to the unconstrained nature of the replacement phrases chosen, it is likely to introduce errors which were not already present. For example, in example 3, it makes changes \emph{racially homogeneous communities} to \emph{homogeneous communities racially}. The adjective \emph{racially} is supposed to qualify the word \emph{homogeneous}, but the output of the algorithm with Distance Function A makes a modification to this, rendering the sentence incorrect. Such errors might creep in due to the lack of constraints on ordering in this function. \\\\
All three distance functions have the Levenshtein Distance and Synset Distance based metrics in common. As a result, they are able to correct the remaining categories of errors. For example, 
\begin{itemize}
\item Substitution based errors, like in example 4, \emph{other protest}$\rightarrow$\emph{others protest}
\item Spelling errors, like in example 2, \emph{precsely}$\rightarrow$\emph{precisely}
\item Missing word errors, like in example 6, \emph{clear the racial history}$\rightarrow$\emph{clear in the racial history}
\end{itemize}
The downside of this method is that it is highly dependent on the corpus being used, and is an extremely domain-dependent algorithm. The words used in the sentences that are being corrected must bear a high degree of correlation to the words used in the corpus. Otherwise, just in order to increase the likelihood of a sentence according to the corpus, some phrases might be substituted by completely unrelated phrases. Also, how this technique will perform on an extremely free word order language like Sanskrit is unknown, largely because of the absence of a suitable corpus in such a language and our lack of comprehensibility of the language. Thirdly, the corpus used must be large enough for a good enough correction. A small corpus will either miss out on possible sources of error introduce new errors just to make the sentence more likely according to the corpus.\\\\
This algorithm can find application in a number of areas. For example, we can use this for correction of sentences output by machine translation systems. If we have a small bitext, but a large enough monolingual corpus in the target language, we can use this algorithm with the monolingual corpus for postprocessing of the translation outputs. This algorithm can also be used for domain specific correction of machine translation outputs. Suppose we have a bitext that is generic (not adhering to a particular domain), but a domain specific corpus in the target language. Once the machine translation system generates the outputs, we can filter out the domain specific sentences and use this algorithm on the monolingual corpus to correct those sentences that belong to the particular domain. This method can also be used for second language learning, by correcting the sentences generated by non-native speakers of a particular language. 
\chapter{Conclusion}
In this work, we present two algorithms to correct noisy sentences. One performs reasonably well, whereas the other fails completely. The fixed length phrase based algorithm tries to make too many corrections to the original text and in the process actually increases the number of errors in the text. This problem is rectified by the Dynamic Programming based approach. This approach actually formulates the problem as a decoding task and uses a model similar to phrase based decoding models in order to make corrections to the text. This model, as seen in Chapter 4, is able to correct a number of errors and increase the fluency of the sentences according to the language model used. We have also suggested a number of applications of this algorithm, like correcting machine translation output and domain specific correction of machine translation output. And as an example, we have explored the machine translation output correction application. \\\\
The primary contribution of the thesis is a method that is language independent and relatively simple. The number of methods explored previously by researchers has the disadvantage of being either language dependent or highly computation intensive. The Dynamic Programming based method is neither very much computation intensive, nor is it at all language dependent. Besides, it has quite low time and space complexities. The thing that distinguishes it from other approaches tried previously is the likeness of the task to decoding. As far as our knowledge goes, nobody has tried to apply decoding algorithms to correct errors in sentences previously. Another major contribution of this work is the varying set of distance functions. The functions cover almost all types of errors that can possibly occur and can be computationally modeled feasibly. \\\\
Much can be done to improve the performance of this algorithm and to build up on the work done. Firstly, we have applied this on a domain specific corpus. If a corpus can be constructed which incorporates data from all domain, the results might be interesting to investigate. Whether this algorithm can work with data from different domains taken together remains to be seen. Another direction that may be followed by future researchers is investigating how this algorithm works on different classes of languages and whether the type of language has any bearing upon the distance function being used. For example, it is likely that for a highly free word order language like Sanskrit, the Distance Function A might be the best choice because it does not constrict the word ordering by any means. A third approach may be to augment the set of Distance Functions in order to cover more errors. How the algorithm performs under these conditions remains to be seen. 
\bibliographystyle{unsrt}
\bibliography{thesis_06CS3031}

\begin{thebibliography}{10}

\bibitem{BRAINMODEL}
M.~Breakspear, V.~Jirsa, and G.~Deco.
\newblock Computational models of the brain: from structure to function.
\newblock {\em Neuroimage}, 3(52), 2010.

\bibitem{StatNLP}
Christopher~D. Manning and Hinrich Schutze.
\newblock {\em Foundations of Sttistical Natural Language Processing}.
\newblock MIT Press, 1 edition, May 1999.

\bibitem{Backoff}
S.~M. Katz.
\newblock Estimation of probabilities from sparse data for the language model
  component of a speech recognizer.
\newblock {\em IEEE Trans.Acoust., Speech, SignaI Processing}.

\bibitem{smoothing}
Stanley~F. Chen and Joshua Goodman.
\newblock An empirical study of smoothing techniques for language modeling.
\newblock {\em Computer Speech {\&} Language}, 13(4):359--393, 1999.

\bibitem{NCM}
Peter~F. Brown, Stephen~Della Pietra, Vincent J.~Della Pietra, and Robert~L.
  Mercer.
\newblock The mathematic of statistical machine translation: Parameter
  estimation.
\newblock {\em Computational Linguistics}, 19(2):263--311, 1993.

\bibitem{ParallelCorpus}
B.~Harris.
\newblock Bi-text, a new concept in translation theory.
\newblock {\em Language Monthly(UK)}.

\bibitem{MERT}
Franz~Josef Och.
\newblock Minimum error rate training in statistical machine translation.
\newblock In {\em ACL}, pages 160--167, 2003.

\bibitem{Koehn}
Philipp Koehn, Franz~Josef Och, and Daniel Marcu.
\newblock Statistical phrase-based translation.
\newblock In {\em HLT-NAACL}, 2003.

\bibitem{Greedydec}
Ulrich Germann.
\newblock Greedy decoding for statistical machine translation in almost linear
  time.
\newblock In {\em HLT-NAACL}, 2003.

\bibitem{pharaoh}
Philipp Koehn.
\newblock Pharaoh: A beam search decoder for phrase-based statistical machine
  translation models.
\newblock In {\em AMTA}, pages 115--124, 2004.

\bibitem{WordGraphdec}
N.~Ueffing, F.~J. Och, and H.~Ney.
\newblock Generation of word graphs in statistical machine translation.
\newblock pages 156--163, Philadelphia, PA, July 2002.

\bibitem{stringtotree}
Kenji Yamada and Kevin Knight.
\newblock A decoder for syntax-based statistical mt.
\newblock In {\em ACL}, pages 303--310, 2002.

\bibitem{CubePrune}
David Chiang.
\newblock Hierarchical phrase based translation.
\newblock {\em Computational Linguistics}, 2007.

\bibitem{truscott}
The effect of error correction on learners' ability to write accurately.
\newblock {\em Journal of Second Language Writing}, 16(4):255 -- 272, 2007.

\bibitem{Kukich}
Karen Kukich.
\newblock Techniques for automatically correcting words in text.
\newblock {\em ACM Comput. Surv.}, 24:377--439, December 1992.

\bibitem{Zobel}
Justin Zobel and Philip Dart.
\newblock Finding approximate matches in large lexicons, 1995.

\bibitem{trinago}
Fran\c{c}ois de~Bertrand~de Beuvron and Philippe Trigano.
\newblock Hierarchically coded lexicon with variants.
\newblock {\em IJPRAI}, 9(1):145--165, 1995.

\bibitem{oflazer}
Kemal Oflazer.
\newblock Error-tolerant finite-state recognition with applications to
  morphological analysis and spelling correction.
\newblock {\em Computational Linguistics}, 22(1):73--89, 1996.

\bibitem{wagner}
Robert~A. Wagner and Michael~J. Fischer.
\newblock The string-to-string correction problem.
\newblock {\em J. ACM}, 21(1):168--173, 1974.

\bibitem{schultzmihov}
Klaus~U. Schulz and Stoyan Mihov.
\newblock Fast string correction with levenshtein automata.
\newblock {\em IJDAR}, 5(1):67--85, 2002.

\bibitem{levenshtein}
Vladimir~I. Levenshtein.
\newblock {Binary codes capable of correcting deletions, insertions, and
  reversals}.
\newblock Technical Report~8, 1966.

\bibitem{Hassan}
Ahmed Hassan, Sara Noeman, and Hany Hassan.
\newblock Language independent text correction using finite state automata,
  2008.

\bibitem{rothgoldberg}
Andrew~R. Golding and Dan Roth.
\newblock Applying winnow to context-sensitive spelling correction.
\newblock {\em CoRR}, cmp-lg/9607024, 1996.

\bibitem{brillmangu}
Lidia Mangu and Eric Brill.
\newblock Automatic rule acquisition for spelling correction.
\newblock In {\em ICML}, pages 187--194, 1997.

\bibitem{knightchander}
K.~Knight and I.~Chander.
\newblock {\em Proceedings of the Twelfth National Conference on Artificial
  Intelligense}, pages 779--784, 1994.

\bibitem{stymne}
Sara Stymne.
\newblock A comparison of merging strategies for translation of german
  compounds.
\newblock In {\em EACL (Student Research Workshop)'09}, pages 61--69, 2009.

\bibitem{elming}
Jakob Elming.
\newblock Transformation-based correction of rule-based mt.
\newblock In {\em 11th Annual Conference of the European Association for
  Machine Translation}.

\bibitem{stymne2}
Sara Stymne and Lars Ahrenberg.
\newblock Using a grammar checker for evaluation and postprocessing of
  statistical machine translation.
\newblock In {\em LREC'10}, pages --1--1, 2010.

\bibitem{xia}
Fei Xia and Michael McCord.
\newblock Improving a statistical mt system with automatically learned rewrite
  patterns.
\newblock In {\em Proceedings of the 20th international conference on
  Computational Linguistics}, COLING '04, Stroudsburg, PA, USA, 2004.
  Association for Computational Linguistics.

\bibitem{niessenney}
Sonja Niessen and Hermann Ney.
\newblock Statistical machine translation with scarce resources using
  morpho-syntactic information.
\newblock {\em Comput. Linguist.}, 30:181--204, June 2004.

\bibitem{collins}
Michael Collins, Philipp Koehn, and Ivona Kucerova.
\newblock Clause restructuring for statistical machine translation.
\newblock In {\em ACL}, 2005.

\bibitem{wang}
Chao Wang, Michael Collins, and Philipp Koehn.
\newblock Chinese syntactic reordering for statistical machine translation.
\newblock In {\em EMNLP-CoNLL}, pages 737--745, 2007.

\bibitem{gammon}
Michael Gamon, Claudia Leacock, Chris Brockett, William~B. Dolan, Jianfeng Gao,
  Dmitriy Belenko, and Alexandre Klementiev.
\newblock Using statistical techniques and web search to correct esl errors.
\newblock {\em CALICO Journal}, 26(3):491--511, 2009.

\bibitem{schierle}
Martin Schierle, Sascha Schulz, and Markus Ackermann.
\newblock From spelling correction to text cleaning - using context
  information.
\newblock In {\em GfKl}, pages 397--404, 2007.

\bibitem{jurafsky}
Daniel Jurafsky and James~H. Martin.
\newblock {\em Speech and Language Processing (2nd Edition) (Prentice Hall
  Series in Artificial Intelligence)}.
\newblock Prentice Hall, 2 edition, 2008.

\bibitem{BLEU}
Kishore Papineni, Salim Roukos, Todd Ward, and Wei-Jing Zhu.
\newblock Bleu: a method for automatic evaluation of machine translation.
\newblock In {\em ACL}, pages 311--318, 2002.

\bibitem{error}
Saadiyah Darus and Kaladevi Subramaniam.
\newblock Error analysis of the written english essays of secondary school
  students in malaysia: A case study, 2009.

\bibitem{GIZA}
Franz~Josef Och and Hermann Ney.
\newblock A systematic comparison of various statistical alignment models.
\newblock {\em Computational Linguistics}, 29(1):19--51, 2003.

\bibitem{SRILM}
A.~Stolcke.
\newblock Srilm -- an extensible language modeling toolkit.
\newblock volume~2, pages 901--904, Denver, 2002.

\bibitem{moses}
Philipp Koehn, Hieu Hoang, Alexandra Birch, Chris Callison-Burch, Marcello
  Federico, Nicola Bertoldi, Brooke Cowan, Wade Shen, Christine Moran, Richard
  Zens, Chris Dyer, Ondrej Bojar, Alexandra Constantin, and Evan Herbst.
\newblock Moses: Open source toolkit for statistical machine translation.
\newblock In {\em ACL}, 2007.

\bibitem{EUROPARL}
Philipp Koehn.
\newblock Europarl: A parallel corpus for statistical machine translation.
\newblock MT Summit.

\bibitem{cyk1}
John Cocke.
\newblock {\em Programming languages and their compilers: Preliminary notes}.
\newblock Courant Institute of Mathematical Sciences, New York University,
  1969.

\bibitem{cyk2}
Daniel~H. Younger.
\newblock {Recognition and parsing of context-free languages in time $n^3$}.
\newblock {\em Information and Control}, 10(2):189--208, 1967.

\bibitem{cyk3}
T.~Kasami.
\newblock An efficient recognition and syntax-analysis algorithm for
  context-free languages.
\newblock {\em Scientific report AFCRL-65-758}.

\bibitem{wordnet}
George~A. Miller.
\newblock Wordnet: A lexical database for english.
\newblock {\em Commun. ACM}, 38(11):39--41, 1995.

\end{thebibliography}
\addcontentsline{toc}{chapter}{Bibliography}
\end{document}